\documentclass[aps,pra,onecolumn,superscriptaddress,groupedaddress,nofootinbib]{revtex4} 

\usepackage{natbib}
\usepackage{amssymb,amsmath,latexsym}
\usepackage{amsthm}
\usepackage{hyperref}
\usepackage{bm}
\usepackage{microtype}
\usepackage{url}
\usepackage{graphicx}
\usepackage{listings}
\usepackage{xspace}
\usepackage[T1]{fontenc}
\usepackage[utf8]{inputenc}

\newcommand{\dropcap}[1]{#1}

\usepackage{xcolor}
\theoremstyle{plain}

\theoremstyle{definition}

\theoremstyle{remark}

\usepackage{xargs}
\usepackage[colorinlistoftodos,prependcaption]{todonotes}
\newcommandx{\unsure}[2][1=]{\todo[linecolor=red,backgroundcolor=red!25,bordercolor=red,#1]{#2}}
\newcommandx{\change}[2][1=]{\todo[linecolor=blue,backgroundcolor=blue!25,bordercolor=blue,#1]{#2}}
\newcommandx{\note}[2][1=]{\todo[linecolor=green,backgroundcolor=green!25,bordercolor=green,#1]{#2}}
\newcommandx{\improvement}[2][1=]{\todo[linecolor=purple,backgroundcolor=purple!25,bordercolor=purple,#1]{#2}}
\newcommandx{\thiswillnotshow}[2][1=]{\todo[disable,#1]{#2}}

\graphicspath{{./figs/}}

\usepackage{tikz}
\usetikzlibrary{shapes.geometric, arrows, fit, calc}

\usepackage{enumitem}
\setlist[itemize]{leftmargin=*}

\definecolor{codegreen}{rgb}{0,0.6,0}
\definecolor{codegray}{rgb}{0.5,0.5,0.5}
\definecolor{codepurple}{rgb}{0.58,0,0.82}
\definecolor{backcolour}{rgb}{0.95,0.95,0.92}

\lstdefinestyle{python}{
    commentstyle=\color{codegreen},
    keywordstyle=\color{magenta},
    numberstyle=\tiny\color{codegray},
    stringstyle=\color{codepurple},
    basicstyle=\ttfamily\footnotesize,
    breakatwhitespace=false,         
    breaklines=true,                 
    captionpos=b,                    
    keepspaces=true,                 
    numbers=left,                    
    numbersep=5pt,                  
    showspaces=false,                
    showstringspaces=false,
    showtabs=false,                  
    tabsize=2
}

\lstdefinestyle{stdout}{
    commentstyle=\color{codegreen},
    keywordstyle=\color{magenta},
    numberstyle=\tiny\color{codegray},
    stringstyle=\color{codepurple},
    basicstyle=\ttfamily\footnotesize,
    breakatwhitespace=false,         
    breaklines=true,                 
    captionpos=b,                    
    keepspaces=true,                 
    showspaces=false,                
    showstringspaces=false,
    showtabs=false,                  
    tabsize=2
}

\lstnewenvironment{pycode}{\lstset{style=python,language=Python}}{}
\lstnewenvironment{pyout}{\lstset{style=python,language=Python}}{}
\lstnewenvironment{pyoutbr}{\lstset{style=python,language=Python}}{}
\lstnewenvironment{stdout}{\lstset{style=stdout}}{}

\newcommand{\py}[1]{\lstinline[language=Python]{#1}}
\newcommand{\qualg}{\py{QuAlg}\xspace}
\usepackage[most]{tcolorbox}
\usepackage{etoolbox}
\tcbset{
}
\BeforeBeginEnvironment{pycode}{\begin{tcolorbox}[colback=white]}%
\AfterEndEnvironment{pycode}{\end{tcolorbox}}%
\BeforeBeginEnvironment{pyout}{\begin{tcolorbox}[colback=white]}%
\BeforeBeginEnvironment{pyoutbr}{\begin{tcolorbox}[breakable,colback=white]}%
\AfterEndEnvironment{pyout}{\begin{tcolorbox}[title=Output]}%
\AfterEndEnvironment{pyoutbr}{\begin{tcolorbox}[title=Output]}%
\AfterEndEnvironment{stdout}{\end{tcolorbox}\end{tcolorbox}}%

\begin{document}

\title{QuAlg - A symbolic algebra package for quantum information}
\author{Axel Dahlberg}
\affiliation{QuTech, Lorentzweg 1, 2628 CJ Delft, Netherlands}
\email[]{e.a.dahlberg@tudelft.nl}

\begin{abstract}
    We introduce QuAlg, an open-source symbolic algebra package for quantum information.
    QuAlg supports working with qubit-states, qudit-states, fock-states and even wave-functions in infinite dimenional Hilbert spaces.
    States can have amplitudes which are pure complex numbers or more complicated symbolic expressions involving for example delta-distributions.
    These symbolic expressions can be automatically simplified and integrated.
    Furthermore, arbitrary operators can be defined, even transformations between different Hilbert spaces.
    Operators can be applied to states or used to defined measurement operators.
\end{abstract}

\maketitle

\section{Introduction}
\dropcap{T}here are many packages and tools for representing and simulating quantum states and operations, such as netsquid~\cite{netsquid}, qutip~\cite{Johansson2012qutip}, Qiskit~\cite{qiskit} and many more\footnote{See \url{https://quantiki.org/wiki/list-qc-simulators} for a longer list.}.
Most of these packages are numerical and for example represent quantum states as vectors or matrices of complex numbers.
However, in some cases it is convenient to be able to express amplitudes, states and operators symbolically.
For example, in some cases, even though the symbolic expression is small, the equivalent numerical expression might be very large, or even infinite-dimensional.

There already exist some projects for providing symbolic algebra for quantum information, such as QNET\footnote{Which recently seem to have been forked as QAlgebra~\cite{qalgebra}.}~\cite{qnet}, Sym$\Psi$~\cite{sympsi}, QuantumComputing (Mathematica package)~\cite{quantumcomputingnb}.
To add to this list, we have now developed \qualg.
One of the features that makes \qualg unique, is the ability to express infinite-dimensional states and operators, using creation and annihilation operators.
This is clearly impossible using numerical calculations.
Amplitudes of such states often involve complex integrals of symbolic functions and delta-distributions which \qualg can evalute and simplify.

The development of \qualg started with a specific use-case in mind, which revolved around computing effective POVM-elements~\cite{dahlberg2019linklayer} of interfering photons with non-unit visibility in a beam-splitter measurement.
This was originally done by hand, which involved evaluating many integrals with delta-distributions.
The original calculation can be found in the appendix of \cite{dahlberg2019linklayer}.
However, in later efforts we wanted to model the interference of wave-packets with higher photon-counts.
This would involve computing the elements of 28 POVM-elements, each being a 16-by-16 matrix, which elements would be found by evaluating large symbolic infinite-dimensional integrals.
Thus, \qualg was born.

\qualg is freely available on \href{https://github.com/AckslD/QuAlg}{github} and \href{https://pypi.org/project/qualg/}{pypi} and can be installed by doing
\begin{pycode}
pip3 install qualg
\end{pycode}

In this note we introduce \qualg.
In section~\ref{sec:framework} we describe the framework of \qualg and how states and operators are represented.
How to use \qualg and its tools is described in section~\ref{sec:usage} by showing a few small examples.
Section~\ref{sec:povm_example} contains a larger example which, using \qualg, computes the same POVM-elements that were computed by hand in appendix of~\cite{dahlberg2019linklayer}

\section{Framework}\label{sec:framework}
\subsection{States}
\subsubsection{BaseState}
The core object in \qualg is the \emph{base-state}.
A certain class of state, such as qubit-states, qudit-states or fock-states, is implemented by inheriting from \py{BaseState}.
An implementation of \py{BaseState} defines how a subset of the states, usually this is a basis of the Hilbert space, act under for example inner-product and tensor-product.
For example, the built-in \py{QubitBaseState} represent \emph{qubit-base-states} as a string of \py{'0'}s and \py{'1'}s.
General states are then built up by representing linear combinations of these base-states, see section~\ref{sec:gen_states} below.

Any implementing of \py{BaseState} must define the following methods:
\begin{itemize}
    \item \py{__eq__}: Specifies when base-states are considered equal and therefore when their amplitudes in a linear combination can be combined.
        For example \py{QubitBaseState}s are simply considered equal if their string of \py{'0'}s and \py{'1'}s are equal.
    \item \py{inner_product}: Specifies how to take inner-product of base-states.
        When the base-states form a basis, such as \py{QubitBaseState}, this function returns \py{1} if the states are equal and otherwise \py{0}.
        Base-states, such as \py{FockBaseState}, might on the other hand return a scalar containing a delta-distribution.
    \item \py{tensor_product}: Specifies how to take tensor-product of base-states.
        For example for \py{QubitBaseState}, tensor-product is simply done by concatenating the strings of \py{'0'}s and \py{'1'}s.
    \item \py{_compatible}: Used to check if two base-states are compatible for addition.
        For example for \py{QubitBaseState} this is whether they represent states of equal number of qubits.
    \item \py{__str__}: Returns a string representing the state as seen as a ket, e.g. \py{'|010>'}.
    \item \py{_bra_str}: Returns a string representing the state as seen as a bra, e.g. \py{'<010|'}.
\end{itemize}

\subsubsection{General states}\label{sec:gen_states}
A general state is represented as a linear combination of base-states, using the class \py{State}.
The class \py{State} is agnostic to the base-state implementation and simply stores a state as a hash map from base-states to scalars.
Scalars can be simply complex numbers or also more complicated symbolic expressions involving delta-distributions or variables, see section~\ref{sec:variables}.
Each non-zero term of a state is thus an entry in this hash map, which values are the amplitudes of the terms.
For example, the state $\frac{1}{2}(|00\rangle + |11\rangle)$ could be represented as
\begin{pycode}
{
    QubitBaseState('00'): 1 / np.sqrt(2),
    QubitBaseState('11'): 1 / np.sqrt(2),
}
\end{pycode}
Operations are then done by updating this hash map.
For example, addition is done by adding the values corresponding to equal keys.
Inner-product and tensor-product is done by taking the Cartesian product and using the \py{inner_product} or \py{tensor_product} defined by the base-state on each term.

\subsection{Operators}
Operators are handled similarly to states.
There is a \py{BaseOperator} representing a single term of an operator by holding two base-states: \py{left} and \py{right}, which form the operator $|\mathrm{left}\rangle\langle\mathrm{right}|$.
In the same way as general states, general operators are stored as hash maps from \py{BaseOperator}s to scalars.
We note that the \py{left} and \py{right} base-states of the \py{BaseOperator} do not need to live in the same Hilbert space.
In this way, operators can represent mappings from for example qubit-states to fock-states and in particular isometries between different Hilbert spaces.
However, the \py{left} (\py{right}) base-states of each term in an operator must be compatible for addition, as defined by the method \py{_compatible}.

\section{Usage}\label{sec:usage}
In this section we go over how to make use of the core concepts in \qualg by showing small examples.
Possible output from \py{print}-statements are shown in a white box below the code snippet.
A code-snippet might depend on state formed by running previous code-snippets to not have to duplicate code.

\subsection{Variables}\label{sec:variables}
One of the key points of \qualg is that variables can be used instead of numbers to represent for example quantum states or operators.
Let's first look at how one can work directly with variables in \qualg.

A variable can be created as follows:
\begin{pyout}
from qualg.scalars import Variable

# A variable 'a'
a = Variable('a')
print(a)
\end{pyout}
\begin{stdout}
a
\end{stdout}
A variable is assumed to be a complex number so it can be conjugated as follows
\begin{pyout}
print(a.conjugate())
\end{pyout}
\begin{stdout}
(a*)
\end{stdout}
\qualg will also recognise certain expression, such as when taking the product of something with its own conjugate:
\begin{pyout}
expr = a * a.conjugate()
print(expr)
\end{pyout}
\begin{stdout}
|a|^2
\end{stdout}
One can easily create arithmetic expressions using the standard addition and multiplication operators:
\begin{pyout}
# Create four variables
a, b, c, d = (
    Variable(name) for name in ['a', 'b', 'c', 'd']
)
expr = (a + b) * (c + d)
print(expr)
\end{pyout}
\begin{stdout}
(a + b)*(c + d)
\end{stdout}
Expressions can be simplified, using \py{simplify}, or expanded, using \py{expand}, to only have sums of products:
\begin{pyout}
from qualg.toolbox import simplify, expand
expr = expand(expr)
print("Expanded expression:")
print(expr)

expr = simplify(expr)
print("Simplified expanded expression:")
print(expr)
\end{pyout}
\begin{stdout}
Expanded expression:
(0 + 0*c + 0*d + 0*a + a*c + a*d + 0*b + b*c + b*d)
Simplified expanded expression:
(a*c + a*d + b*c + b*d)
\end{stdout}

\subsection{States}
In this section we go over how to define and work with quantum states using \qualg.
\subsubsection{Single-qubit states}
We first look at how one can create qubit states.
A \py{BaseQubitState} (subclass of \py{BaseState}) represents a single term in a quantum state and can be used to build a basis for such.
\begin{pyout}
import numpy as np
from qualg.q_state import BaseQubitState

# Create a basis for single-qubit states
bs = [BaseQubitState(i) for i in '01']
print(bs[0])
print(bs[1])
\end{pyout}
\begin{stdout}
|0>
|1>
\end{stdout}
Using these base states we can construct arbitary states.
To do so we first need to make these base states actual states, which we can then add together to construct superpositions of such.
\begin{pycode}
s0 = bs[0].to_state()
s1 = bs[1].to_state()
\end{pycode}
Let's now for example construct the state $|+\rangle=\frac{1}{\sqrt{2}}(|0\rangle + |1\rangle)$:
\begin{pyout}
h0 = (s0 + s1) * (1 / np.sqrt(2))
print(h0)
\end{pyout}
\begin{stdout}
0.7071067811865475*|0> + 0.7071067811865475*|1>
\end{stdout}
One thing you will see when working with \qualg is that all objects have a nice string-representation and can be printed.
This is one of the aims of \qualg, to allow a user to inspect the expressions formed by calculations.

Variables, as we have seen above, can also be used to represent states symbolically:
\begin{pyout}
a = Variable('a')
b = Variable('b')
s_psi = (a * s0 + b * s1)
print(s_psi)
\end{pyout}
\begin{stdout}
a*|0> + b*|1>
\end{stdout}
The norm of this state, that is the inner-product with itself, can then be computed as:
\begin{pyout}
inner = s_psi.inner_product(s_psi)
print(inner)
\end{pyout}
\begin{stdout}
(|a|^2 + |b|^2)
\end{stdout}

\subsubsection{Multi-qudit states}
Let's say that we instead would like to work with states on two qutrits.
We can easily do this as follows:
\begin{pyout}
from itertools import product
from qualg.states import State
from qualg.q_state import BaseQuditState

levels = 3
bs = [
    BaseQuditState(f"{i}{j}", base=levels)
    for i, j in product(range(levels), range(levels))
]
print(bs[0])
print(bs[-1])
\end{pyout}
\begin{stdout}
|00>
|22>
\end{stdout}
By changing \py{levels} in the code snippet above, one can work with qudits of different dimensions.
Similarly to before we can now construct superpositions of these base states:
\begin{pyout}
superpos = sum([b.to_state() for b in bs], State())
# Normalize by dividing by sqrt(levels ** 2)
superpos /= levels
print(superpos)
\end{pyout}
\begin{stdout}
0.3333333333333333*|00> + 0.3333333333333333*|01>...
...+ 0.3333333333333333*|02> + 0.3333333333333333*|10>...
...+ 0.3333333333333333*|11> + 0.3333333333333333*|12>...
...+ 0.3333333333333333*|20> + 0.3333333333333333*|21>...
...+ 0.3333333333333333*|22>
\end{stdout}

\subsubsection{Fock states}
We can also work with states in second quantization, where these are represented as excitations of some given mode.
To do this we first need to define our creation operators:
\begin{pyout}
from qualg.fock_state import (
    FockOp,
    FockOpProduct,
    BaseFockState,
)

aw = FockOp('a', 'w')
print(aw)
\end{pyout}
\begin{stdout}
a+(w)
\end{stdout}
What \py{aw} represents above, is a creation operator in mode $a$, with variable $\omega$, where for example $\omega$ could be the frequency of a wave-packet.
This can then be used to define the state where this operator acts on the vacuum:
\begin{pyout}
s = BaseFockState([aw])
print(s)
\end{pyout}
\begin{stdout}
a+(w)^1|0>
\end{stdout}
Similarly to qubit-states, inner-products can also be computed between fock-states as:
\begin{pyout}
aw = FockOp('a', 'w')
av = FockOp('a', 'v')
saw = BaseFockState([aw])
sav = BaseFockState([av])
print(saw.inner_product(sav))
\end{pyout}
\begin{stdout}
D[w-v]
\end{stdout}
What we find is that the inner-product becomes a delta-distribution.

Let's now look at a more complicated example where we have states which are superpositions over modes defined by a wave-packet.
For example, let's say that we have one state with the wave-packet $\Phi$ and the other with $\Psi$, as follows:
\begin{pyout}
from qualg.scalars import SingleVarFunctionScalar

phi = SingleVarFunctionScalar("phi", 'w')
psi = SingleVarFunctionScalar("psi", "w")
s_phi = phi * saw.to_state()
s_psi = psi * saw.to_state()
print(s_phi)
print(s_psi)
\end{pyout}
\begin{stdout}
phi(w)*a+(w)^1|0>
psi(w)*a+(w)^1|0>
\end{stdout}
Which represents the two states
\begin{equation}
    \Phi(\omega)a^\dagger(\omega)|0\rangle
\end{equation}
and
\begin{equation}
    \Psi(\omega)a^\dagger(\omega)|0\rangle.
\end{equation}
We can now take the inner product of these two states:
\begin{pyout}
inp = s_phi.inner_product(s_psi)
print(inp)
\end{pyout}
\begin{stdout}
phi*(w)*psi(w')*D[w-w']
\end{stdout}
What is then left to do is to integrate out this expression since there are integrals from the states which we left out until now.
By default, the \py{integrate} function will integrate out all varibles in the expression if not otherwise specified.
\begin{pyout}
from qualg.integrate import integrate

integrated_inp = integrate(inp)
print(integrated_inp)
\end{pyout}
\begin{stdout}
<phi|psi>
\end{stdout}
We can see that \qualg realised that this expression is in fact the inner product of the two functions $\Phi$ and $\Psi$.
We can see this more clearly by writing out the representation of the state:
\begin{pyout}
print(repr(integrated_inp))
\end{pyout}
\begin{stdout}
InnerProductFunction('phi', 'psi')
\end{stdout}
So we have found that the inner-product of these two states is exactly the inner-product of the wave-packets describing the states.

\subsection{Operators}
Operators work very much like states, in that there is a \py{BaseOperator}-class which represents the terms of an \py{Operator}.
A \py{BaseOperator} is described by a \py{left} and \py{right} \py{BaseState}, i.e. $|\mathrm{left}\rangle\langle\mathrm{right}|$.

\subsubsection{Single-qubit operators}
Let's start with creating some single-qubit base operators.
\begin{pyout}
from qualg.operators import (
    BaseOperator,
    Operator,
    outer_product,
)

# Create a basis for single-qubit states
bs = [BaseQuditState(f"{i}") for i in range(2)]

bp0 = BaseOperator(left=bs[0], right=bs[0])
print(bp0)
\end{pyout}
\begin{stdout}
Op[|0><0|]
\end{stdout}
We can create an \py{Operator} from a \py{BaseOperator} as;
\begin{pyout}
p0 = bp0.to_operator()
print(p0)
\end{pyout}
\begin{stdout}
1*Op[|0><0|]
\end{stdout}
Let's say we want to create the Hadamard operator, we can do this by adding the correct terms of this operator.
However, we can also do this by making use of the \py{outer_product} function on \py{State}s:
\begin{pyout}
# Create standard and Hadamard basis states
s0 = bs[0].to_state()
s1 = bs[1].to_state()
h0 = (s0 + s1) * (1 / np.sqrt(2))
h1 = (s0 - s1) * (1 / np.sqrt(2))

# Create Hadamard gate
h = outer_product(h0, s0) + outer_product(h1, s1)
print(h)
\end{pyout}
\begin{stdout}
0.7071067811865475*Op[|0><0|] ...
...+ 0.7071067811865475*Op[|1><0|] ...
...+ 0.7071067811865475*Op[|0><1|] ...
...+ -0.7071067811865475*Op[|1><1|]
\end{stdout}
We can now let this operator act on a state:
\begin{pyout}
print(h * s0)
print(h * h0)
\end{pyout}
\begin{stdout}
0.7071067811865475*|0> + 0.7071067811865475*|1>
0.9999999999999998*|0>
\end{stdout}

\subsubsection{General operators}
We can also create operators acting on non-qubit states and even operators taking a state in one Hilbert-space to another.

For example let's define an operator from a qubit- to a fock-state:
\begin{pyout}
# Create standard basis states
bs = [BaseQuditState(f"{i}") for i in range(2)]
s0 = bs[0].to_state()
s1 = bs[1].to_state()

# Create two fock states in different modes
a = FockOp('a', 'w')
b = FockOp('b', 'w')
fs0 = BaseFockState([a]).to_state()
fs1 = BaseFockState([b]).to_state()

# Create an operator from a qubit state to a fock state
op = outer_product(fs0, s0) + outer_product(fs1, s1)
print(op)
\end{pyout}
\begin{stdout}
1*Op[a+(w)^1|0><0|] + 1*Op[b+(w)^1|0><1|]
\end{stdout}

\subsection{Measuring}
We can use what we have learned about states and operators to now also measure states given a set of measurement operators.
We can create general POVM measurements but we will here for simplicity only perform a measurement in the standard single-qubit basis.
However general POVMs work in the same way.

To perform a measurement we first need to define the Kraus operators that define the measurement:
\begin{pycode}
from qualg.measure import measure

# Create single-qubit states
s0 = BaseQubitState("0").to_state()
s1 = BaseQubitState("1").to_state()
# Create Hadamard basis state
h0 = (s0 + s1) * (1 / np.sqrt(2))

# Create a projective measurement
P0 = outer_product(s0, s0)
P1 = outer_product(s1, s1)
kraus_ops = {0: P0, 1: P1}
\end{pycode}
What we pass into the measure-function is a dictionary where the keys are the measurement outcomes/labels and the values the Kraus operators.
Note that there is no check that the operators actually form a valid POVM.

Let's now measure some states and see that we get (run the below examples multiple times to see what happens):
\begin{pyout}
# Measure |0>
meas_res = measure(s0, kraus_ops)
print(meas_res)
\end{pyout}
\begin{stdout}
MeasurementResult(outcome=0, probability=1, ...
...post_meas_state=State([(BaseQubitState('0'), 1.0)]))
\end{stdout}

\begin{pyout}
# Measure |1>
meas_res = measure(s1, kraus_ops)
print(meas_res)
\end{pyout}
\begin{stdout}
MeasurementResult(outcome=1, probability=1, ...
...post_meas_state=State([(BaseQubitState('1'), 1.0)]))
\end{stdout}

\begin{pyout}
# Measure |+>
meas_res = measure(h0, kraus_ops)
print(meas_res)
\end{pyout}
\begin{stdout}
MeasurementResult(outcome=0, ...
...probability=0.4999999999999999, ...
...post_meas_state=State([(BaseQubitState('0'), 1.0)]))
\end{stdout}

\section{POVM-example}\label{sec:povm_example}
This section shows how \qualg can be used to compute certain POVM-elements that were previously computed by hand in the appendix of \cite{dahlberg2019linklayer}.
\begin{pyoutbr}
"""
This example shows how the POVMs used in the paper
https://arxiv.org/abs/1903.09778
can be computed using QuAlg.
"""
import numpy as np
from timeit import default_timer as timer

from qualg.scalars import (
    SingleVarFunctionScalar,
    InnerProductFunction,
    ProductOfScalars,
    SumOfScalars,
    is_number,
)
from qualg.q_state import BaseQubitState
from qualg.fock_state import BaseFockState, FockOp
from qualg.operators import Operator, outer_product
from qualg.toolbox import simplify, replace_var
from qualg.integrate import integrate


def get_fock_states():
    """Construct the fock states (eq. (42)-(45) in
    https://arxiv.org/abs/1903.09778)"""
    bs0 = BaseFockState()
    bsc = BaseFockState([FockOp("c", "w1")])
    bsd = BaseFockState([FockOp("d", "w1")])
    phi = SingleVarFunctionScalar("phi", "w1")
    psi = SingleVarFunctionScalar("psi", "w1")

    s0 = bs0.to_state()
    f = 1 / np.sqrt(2)
    sphi = f * phi * (bsc.to_state() + bsd.to_state())
    spsi = f * psi * (bsc.to_state() - bsd.to_state())

    bscc = BaseFockState([
        FockOp("c", "w1"),
        FockOp("c", "w2"),
    ])
    bscd = BaseFockState([
        FockOp("c", "w1"),
        FockOp("d", "w2"),
    ])
    bsdc = BaseFockState([
        FockOp("d", "w1"),
        FockOp("c", "w2"),
    ])
    bsdd = BaseFockState([
        FockOp("d", "w1"),
        FockOp("d", "w2"),
    ])
    phipsi = (
        SingleVarFunctionScalar("phi", "w1") *
        SingleVarFunctionScalar("psi", "w2")
    )
    sphipsi = (1 / 2) * phipsi * (
        bscc.to_state() +
        bsdc.to_state() -
        bscd.to_state() -
        bsdd.to_state()
    )

    return s0, sphi, spsi, sphipsi


def construct_beam_splitter():
    """
    Construct the beam splitter operation (eq. (40)-(41)
    in https://arxiv.org/abs/1903.09778)
    """
    fock_states = list(get_fock_states())
    for i, state in enumerate(fock_states):
        for old, new in zip(['w1', 'w2'], ['b1', 'b2']):
            state = replace_var(state, old, new)
        fock_states[i] = state
    qubit_states = [
        BaseQubitState(b).to_state()
        for b in ["00", "01", "10", "11"]
    ]

    beam_splitter = sum((
        outer_product(fock_state, qubit_state)
        for fock_state, qubit_state in zip(
            fock_states,
            qubit_states,
        )
    ), Operator())

    return beam_splitter.simplify()


def construct_projector(num_left, num_right):
    """Construct the projector (eq. (46)-(51) in
    https://arxiv.org/abs/1903.09778)"""
    vac = BaseFockState([]).to_state()
    c1 = BaseFockState([FockOp('c', 'p1')]).to_state()
    c2 = BaseFockState([FockOp('c', 'p2')]).to_state()
    d1 = BaseFockState([FockOp('d', 'p1')]).to_state()
    d2 = BaseFockState([FockOp('d', 'p2')]).to_state()

    if (num_left, num_right) == (0, 0):
        # P00
        return outer_product(vac, vac)
    elif (num_left, num_right) == (1, 0):
        # P10
        return outer_product(c1, c1)
    elif (num_left, num_right) == (0, 1):
        # P01
        return outer_product(d1, d1)
    elif (num_left, num_right) == (1, 1):
        # P11
        return outer_product(c1@d2, c1@d2)
    elif (num_left, num_right) == (2, 0):
        # P20
        return outer_product(c1@c2, c1@c2)
    elif (num_left, num_right) == (0, 2):
        # P02
        return outer_product(d1@d2, d1@d2)
    else:
        raise NotImplementedError()


def example_states(no_output=False):
    """Example showing how the states can be constructed
    """
    s0, sphi, spsi, sphipsi = get_fock_states()
    if not no_output:
        print(f"State is: {sphi}\n")
    inner = sphi.inner_product(sphi)
    if not no_output:
        print(f"inner product is: {inner}\n")
    simplify(inner)
    if not no_output:
        print(f"after simplify: {simplify(inner)}\n")
    inner = integrate(inner)
    if not no_output:
        print(f"after integrate: {inner}\n")
    return

    inner = simplify(sphipsi.inner_product(sphipsi))
    if not no_output:
        print(inner)
        print(integrate(inner))


def example_beam_splitter(no_output=False):
    """Example showing how the beam spliter operation
    can be constructed"""
    beam_splitter = construct_beam_splitter()
    if not no_output:
        print(f"Beam splitter:\n{beam_splitter}")


def example_projectors(no_output=False):
    """Example showing how to construct the projectors
    and how they act on a given state."""
    s0, sphi, spsi, sphipsi = get_fock_states()

    s = sphi
    p = construct_projector(1, 0)
    inner = (p * s).inner_product(s)
    if not no_output:
        print(integrate(inner))


def ultimate_example(
    indices,
    no_output=False,
    visibility=1.0,
):
    """
    Example for computing one of the POVMs
    (eq. (82) - (87) of https://arxiv.org/abs/1903.09778)
    """
    def convert_scalars(scalar):
        scalar = integrate(scalar)
        if is_number(scalar):
            return scalar
        sequenced_classes = [
            ProductOfScalars,
            SumOfScalars,
        ]
        for sequenced_class in sequenced_classes:
            if isinstance(scalar, sequenced_class):
                return simplify(sequenced_class([
                    convert_scalars(s) for s in scalar
                ]))
        if isinstance(scalar, InnerProductFunction):
            func_names = set(['phi', 'psi'])
            if set(scalar._func_names) == func_names:
                return visibility
        raise RuntimeError(
            f"unknown scalar {scalar} "
            f"of type {type(scalar)}"
        )

    assert len(indices) == 2

    u = construct_beam_splitter()
    p = construct_projector(*indices)
    m = u.dagger() * p * replace_var(u)
    m = simplify(m)
    if not no_output:
        # print(m)
        print(m.to_numpy_matrix(convert_scalars))


def main(no_output=False):
    # example_states(no_output=no_output)
    # example_beam_splitter(no_output=no_output)
    # example_projectors(no_output=no_output)
    ultimate_example(
        (1, 0),
        no_output=no_output,
        visibility=0.9,
    )


if __name__ == '__main__':
    main()
\end{pyoutbr}
\begin{stdout}
[[0.   0.   0.   0.  ]
 [0.   0.5  0.45 0.  ]
 [0.   0.45 0.5  0.  ]
 [0.   0.   0.   0.  ]]
\end{stdout}

\section{Conclusion}
\qualg started with a specific use-case in mind but has been extended to have a more general purpose.
However, there is still a lot that can be added and extended and we encourage anyone who is interested to contribute.
The goal would be to build a community that can maintain this open-source project.
Examples of features that can added or improved are:
\begin{itemize}
    \item Optimize code executing to allow for faster evaluation of large symbolic expressions.
    \item Extend the simplification of symbolic expressions to be able to handle more cases.
        Perhaps integrate with an existing symbolic expression library such as \py{sympy}~\cite{sympy}.
    \item Improving printing. Print large states in table-format for readability.
\end{itemize}

\section*{Acknowledgments}

We thank David Maier and Sam Morley-Short for feedback and comments.
AD was supported by STW Netherlands, and NWO VIDI grant, and an ERC Starting grant. 

\bibliographystyle{unsrt}
\bibliography{refs}

\begin{thebibliography}{1}

\bibitem{netsquid}
QuTech.
\newblock {NetSQUID}.
\newblock \url{https://netsquid.org/}, 2018.

\bibitem{Johansson2012qutip}
J.~R. Johansson, P.~D. Nation, and Franco Nori.
\newblock {QuTiP: An open-source Python framework for the dynamics of open
  quantum systems}.
\newblock {\em Computer Physics Communications}, 183(8):1760--1772, 2012.

\bibitem{qiskit}
{Qiskit}.
\newblock \url{https://qiskit.org/}, 2020.

\bibitem{qalgebra}
{QAlgebra}.
\newblock \url{https://github.com/QAlgebra/qalgebra}, 2020.

\bibitem{qnet}
{QNET}.
\newblock \url{https://github.com/mabuchilab/QNET}, 2020.

\bibitem{sympsi}
{SymPsi}.
\newblock \url{https://github.com/sympsi/sympsi}, 2020.

\bibitem{quantumcomputingnb}
{QuantumComputing (Mathematica package)}.
\newblock \url{https://github.com/jacobmarks/quantum_computing}, 2020.

\bibitem{dahlberg2019linklayer}
Axel Dahlberg, Matthew Skrzypczyk, Tim Coopmans, Leon Wubben, Filip
  Rozp\k{e}dek, Matteo Pompili, Arian Stolk, Przemys\l{}aw Pawe\l{}czak, Robert
  Knegjens, Julio de~Oliveria~Filho, Ronald Hanson, and Stephanie Wehner.
\newblock A link layer protocol for quantum networks.
\newblock In {\em ACM SIGCOMM 2019 Conference}, SIGCOMM '19, page~15, New York,
  NY, USA, 2019. ACM.

\bibitem{sympy}
{SymPy}.
\newblock \url{https://www.sympy.org/en/index.html}, 2020.

\end{thebibliography}

\end{document}